\documentclass[aps, prl, onecolumn, superscriptaddress
]{revtex4-1}
\newcommand{\gguide}

\usepackage{graphicx}
\usepackage{gensymb}
\usepackage{amsmath}
\usepackage{textcomp}
\usepackage{bm}
\usepackage{color}
\usepackage{soul}
\usepackage{verbatim}

\begin{document}

\title{Imaging with a small number of photons}


\title[]{Imaging with a small number of photons}

\author{Peter~A.~Morris}
\email[Email: ]{p.morris.1@research.gla.ac.uk}
\author{Reuben~S.~Aspden}
\email[Email: ]{r.aspden.1@research.gla.ac.uk}
\author{Jessica Bell}
\affiliation{School of Physics and Astronomy, University of Glasgow, Glasgow, G12 8QQ, UK}
\author{Robert~W.~Boyd}
\affiliation{Department of Physics, University of Ottawa, Ottawa, Ontario, Canada}
\affiliation{The Institute of Optics and Department of Physics and Astronomy, University of Rochester, Rochester, NY 14627, USA}
\author{\\Miles~J.~Padgett}
\affiliation{School of Physics and Astronomy, University of Glasgow, Glasgow, G12 8QQ, UK}

\date{\today}


\begin{abstract}
	
Low-light-level imaging techniques have application in many diverse fields, ranging from biological sciences to security. We demonstrate a single-photon imaging system based on a time-gated intensified CCD (ICCD) camera in which the image of an object can be inferred from very few detected photons.  We show that a ghost-imaging configuration, where the image is obtained from photons that have never interacted with the object, is a useful approach for obtaining images with high signal-to-noise ratios. The use of heralded single-photons ensures that the background counts can be virtually eliminated from the recorded images.  By applying techniques of compressed sensing and associated image reconstruction, we obtain high-quality images of the object from raw data comprised of fewer than one detected photon per image pixel.

\end{abstract}

\maketitle

\section{Introduction}

The photons generated through the spontaneous parametric down-conversion (SPDC) process have served as an illumination source for many low-light level applications \cite{strekalov1995, walborn2006, dixon2012}. The SPDC process provides an easily manipulated source of photon pairs with strong correlations in the spatial degrees of freedom of the photons \cite{walborn2010}. Furthermore, the photons can be separated using a beam splitter into two different optical paths or arms of an experiment. These correlations have been exploited in several single-photon imaging experiments, including quantum ghost imaging \cite{pittman1995} and quantum interference imaging \cite{lemos2014}. One method for utilising these correlations is an imaging system where the detection of one of the photons in the generated photon pair is used to herald the arrival of its partner. In such systems, the heralding detector is a large area, single-pixel detector whilst the other, the imaging detector, is spatially resolving. One has two options for object placement- either place the object in the same arm of the experiment as the imaging detector as per a standard imaging system, or, by exploiting the spatial correlations between the two photons, place the object in the heralding detector arm, as demonstrated by Pittman et al.~\cite{pittman1995} in a display of quantum ghost imaging.  Despite the use of a SPDC source, one should note that correlations within a single measurement basis (in this case the position basis) are not in themselves proof of entanglement but rather a utilisation of entanglement \cite{gatti2003, bennink2004}.

Traditionally, within a quantum ghost imaging system, the spatially-resolving detector has been a scanning single-pixel detector. However, basing the system upon a single scanning detector fundamentally limits the detection efficiency to $1/N$, where $N$ is the number of pixels in the image. Overcoming this limitation by using a detector array to increase the detection efficiency enables the acquisition of images whilst illuminating the sample with $N$-times fewer photons. This reduction in the required illumination flux is potentially beneficial for applications in biological imaging, where bleaching or sample damage can occur from a high photon-flux, and also in security, where reducing the photon-flux can make the system covert. Indeed, there are a number of recent papers using detector arrays with single-photon sensitivity \cite{edgar2012, brida2010, fickler2013, aspden2013}.

We characterise our camera-enabled, time-gated imaging system for two different system configurations, either with the object in the heralding arm, as per ghost imaging, or the object in the camera arm of the system. In order to utilise the low-photon flux capabilities of our system, reconstruction techniques are applied to our data that allow us to obtain images using undersampled data sets consisting of fewer than 1 photon per image pixel.  We achieve this by operating within the constraints of Poissonian statistics and exploiting the sparsity of our data when expressed in an appropriate domain to subjectively improve the quality of the reconstructed images. With optimisation we are able to obtain images of our biological sample using fewer detected photons than there are pixels in the image.

\section{Experimental methods}  

\begin{figure}
\begin{center}
\includegraphics[width=\linewidth]{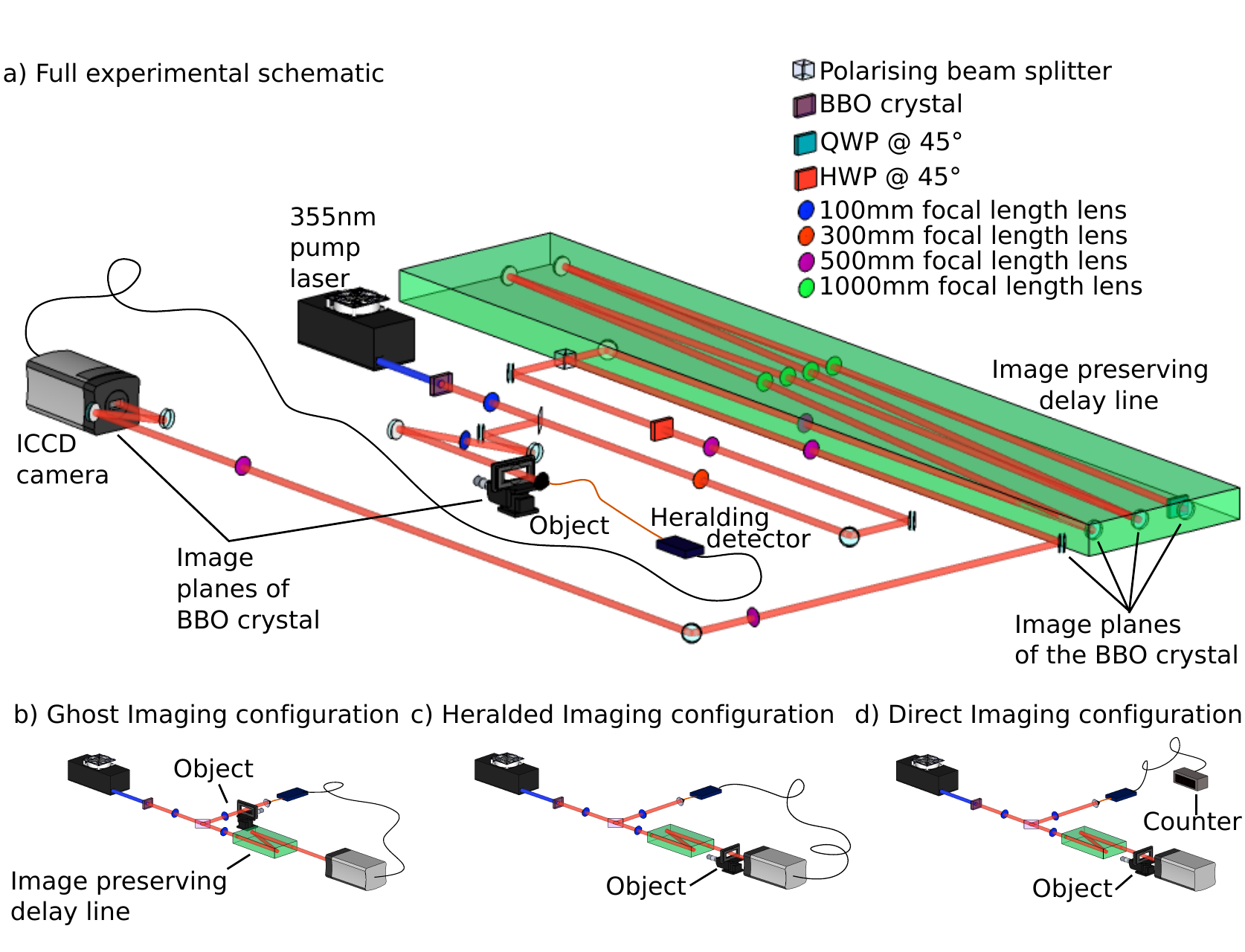}
\caption{a) Full schematic of our imaging system. A $355$~nm laser pumps a BBO crystal to produce collinear down-converted photon pairs at $710$ nm. The output facet of the crystal is imaged onto the plane of the microscope slide (containing our object) and the ICCD camera. The image-preserving delay line is necessary to compensate for the electronic delays in the triggering mechanism. b)-d) Simplified schematics of each imaging configuration. b) Ghost imaging configuration: the object is placed in the heralding arm and the camera is triggered by each photon detection at the heralding detector. c) Heralded imaging configuration: the object is placed in the camera arm and the camera is again triggered by each photon detection at the heralding detector. d) Direct imaging configuration: the object is placed in the camera arm but the camera is triggered by an internal trigger mechanism, with the same trigger rate as number of singles detected at the counter in the heralding arm.}
\label{FIG:setup}
\end{center}
\end{figure}

Our imaging system is similar to that reported in \cite{aspden2013, tasca2013}. We use correlated photons generated by SPDC and a multipixel ICCD triggered by a single-photon avalanche detector (SPAD), the latter acting as the heralding detector. The source of our down-converted photons is a 3-mm-long, non-linear $\beta$-barium borate (BBO) crystal, cut for type-I phase matching and pumped by a horizontally polarised, quasi continuous-wave laser at 355~nm. The laser output is spatially filtered and recollimated to produce a $\approx$1.2~mm (full width half-maximum) fundamental Gaussian beam at the input facet of the down-conversion crystal. The generated near-collinear beam of frequency-degenerate, down-converted photons is selected through the use of high-transmission interference filters with a 10~nm bandwidth centred on 710~nm. Due to the large transverse Gaussian profile of the pump beam and short length of the down-conversion crystal, our down-converted photons exhibit strong correlations over a wide range of spatial modes \cite{shapiro2012}. Our pairs of correlated photons are separated using a pellicle beam splitter (BS) that directs the separated photons into the camera arm and the heralding arm of the system. Each arm has a magnification of M=3 between the plane of the down-conversion crystal and the planes of the object/camera. The object is placed on a microscope slide positioned in the image plane of the crystal in either the heralding or camera arm, depending on the desired system configuration (see figure \ref{FIG:setup}). The camera is also positioned in an image plane of the crystal/object. Our object is thus illuminated by a spatially incoherent, multimode beam with a FWHM of approximately 3.6~mm. The photons in the heralding arm are collected by a detector consisting of an x4 objective lens, a 400 $\mu m$ core multimode fibre and a SPAD. This heralding detector registers the detection of a photon but records no spatial information. 

There are two timing measures of relevance when using an ICCD camera. The first of these is the intensifier gate-width, during which any single input photon is amplified by the intensifier and the event recorded on the CCD chip. This gate-width has a typical duration of several nano-seconds. The second is the CCD exposure time, which is the time between each readout of the CCD chip, typically several seconds. Of course, the intensifier can fire many times during each exposure and thus each frame that is read out is an accumulation of all the detected single-photon events acquired during the exposure time.  

The intensifier of the ICCD camera can be triggered using either an external pulse from the heralding SPAD or by using an internal pulse generator. When triggered using an external pulse, the gate-width of the intensifier is set by the width of the input transistor-transistor logic (TTL) pulse from the SPAD ($\approx15$ ns).  To ensure the photons detected at the heralding detector and at the camera are from the same correlated photon pair, the electronic delay in the ICCD triggering mechanism must be compensated for by the introduction of additional optical path length in the camera arm \cite{aspden2013}. In our system, we compensate for this electronic delay by introducing a 22~m image preserving, free-space delay line. We attenuate the pump beam in order to all but eliminate the probability of generating multiple photon pairs per pump laser pulse, ensuring that we only record one photon per gating of the ICCD camera.



\section{Image acquisition} 

We acquire images using three different system configurations as shown in figure \ref{FIG:setup}. In the ghost imaging (GI) configuration, the object is placed in the heralding arm, and the camera is triggered externally by the signal from the heralding detector. Thus an image of the object is formed on the camera, despite none of the imaged photons having interacted with the object. For the heralded imaging (HI) configuration, the camera is again triggered by the external trigger pulse, but the object is placed in an intermediate imaging plane in the camera arm. The camera is therefore triggered for each detected single photon yet the image consists only of the correlated photons that pass through the object. For comparison, we also show direct imaging (DI), where the camera is triggered using its internal trigger mechanism. In this last configuration the image consists only of the subset of photons that pass through the object and arrive at the camera during the camera trigger window by random chance. These three system configurations are illustrated in figure \ref{FIG:setup}. 

\begin{figure}
\begin{center}
\includegraphics[width=\linewidth]{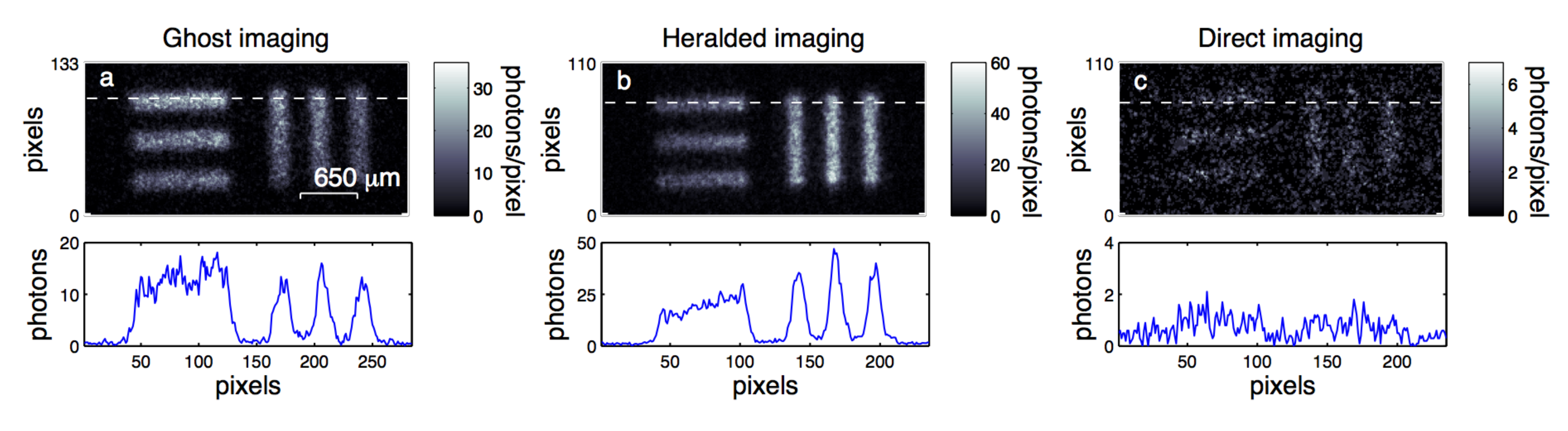}
\caption{Acquired images and associated cross-sections in the different imaging configurations. a) Ghost Imaging (GI) configuration with the object in the heralding arm and the camera triggered by the heralding detector. b)  Heralded Imaging (HI) configuration where the object is in the camera arm and the camera is triggered by the heralding detector and c) Direct Imaging configuration, where the object is in the camera arm and the camera is internally triggered. It can be seen that we obtain a clear image with high contrast in both the GI and HI configurations, whilst the random nature of the detection mechanism in the direct imaging configuration yields only a very low contrast image.}
\label{FIG:results1}
\end{center}
\end{figure}

The images shown in figure \ref{FIG:results1} are formed from the sum of 900 frames each of 2~s exposure, during which time the camera intensifier fires for every trigger pulse received, either from the heralding detector or the internal trigger mechanism. The CCD chip is air cooled to -30$^\circ$C, and we work with a region of interest (ROI) of $600\times600$ pixels, covering an area of $(7.8\times7.8)$~mm$^2$. The exposure time is chosen to ensure that each acquired frame is photon sparse, i.e.~$\ll$1 photon event per pixel \cite{tasca2013}. Photon counting is possible by applying a binary threshold to the value of each pixel in the data read from the ICCD, a fuller description of which is provided in reference \cite{aspden2013}. As part of this photon counting procedure, we calculate a noise probability per pixel by acquiring 100 triggered frames with the camera shutter closed. Plotting a histogram of the output signal from the camera allows us to set a threshold, a signal over which we define as a photon. Using this threshold, we calculate a dark-count probability per pixel arising from the camera read out noise, which we calculate to be $5 \times 10^{-4}$ per frame.

Figure \ref{FIG:results1} shows the images acquired using each of the system configurations. In both the GI and HI configurations, we obtain a clear image of the test target with image contrast of order 40:1.  By comparison, when using the DI configuration, only a very faint image of the object is obtained. The periodic nature of the intensifier trigger in this DI configuration means that the random arrival of the photon and the regular firing of the camera intensifier window only occasionally coincide, and thus the coincidence nature of the system is lost leading to a very low detection efficiency.  Closer inspection of the GI and HI images reveals a slight difference in scale resulting from the magnification in the two arms not being quite the same.  One also notes that although the total number of image photons is similar in the two cases, the GI configuration was obtained with fewer triggers of the intensifier than the HI configuration.  This difference arises because although the photon pair generation rate in the two configurations is the same, when the partially transmitting object is placed in the heralding arm the trigger rate is reduced in proportion to the transmission of the object.  For high flux rates the GI configuration may therefore prove to be advantageous since it makes a lower technical demand upon the ICCD camera.


\section{Optimisation of reconstructed image}

For the imaging system to be applied in ultra-low light conditions, one fundamental question is ``how many photons does it take to form an image?'' Simplistically speaking, one requires many photons per pixel (typically 10,000s photons/pixel for a conventional imaging system), so that the intensity of each pixel is not unduly subject to the Poissonnian statistics associated with the quantisation of the number of individual detected photon \cite{morris1984}.  However, when an image is sparse in a chosen basis, it is possible to implement compressive techniques in order to store or even reconstruct the image from far fewer measurements than this simplistic statement implies \cite{candes2008, romberg2008, donoho2006, shechtman2011}. These reconstruction techniques have also been shown to enhance efficiency in applications requiring the exploration of a large state space, for example in quantum state tomography \cite{shabani2011} and more recently in quantum imaging. This latter use of compressive techniques in a quantum imaging system allowed an image to be reconstructed using single-pixel detectors and far fewer samples than required by the Nyquist limit, albeit whilst still requiring many photons per pixel \cite{zerom2011, magana2013}.

Even for our longer acquisition times, our images have a very small ($<20$) number of detected photons per pixel, and thus, even for a uniform transmittance region of the object, the difference between neighbouring pixels in our images show a large variation inherent in the Poissonian statistics of the shot-noise. Therefore, although the signal to background ratio (SBR) of our images is high, the signal to noise ratio (SNR) is not. However, the noise contributions in our images are well-defined both in terms of the Poissonian characteristics of photon counting and a known rate of noise events. 

Real images are usually sparse in the spatial frequency domain, meaning they contain comparatively few significant spatial frequency components, a concept that forms the basis of JPEG image compression. The concepts of compressed sensing allow us to utilise this sparsity to infer an image from fewer photons than necessary in standard imaging techniques. Here we modify the image data to maximise the sparsity of the contributing spatial frequencies whilst maintaining the likelihood of the resulting image within the bounds set by the Poissonian statistics of the original data.

We denote the measured number of photons for each of the $N$ image pixels to be an integer $n_j$ and the fractional intensity of each pixel of the modified image to be $I_j$ .  Given an estimated dark count rate of $\varepsilon$ per pixel, the Poisson probability distribution of measuring n photons given a pixel intensity of I is

\begin{equation}
P (I_j;n_j) = \frac{(I_j+\varepsilon)^{n_j}  e^{-(I_j+\varepsilon)}}{n_j!}
\end{equation}

from which we can state the log likelihood of a modified image, $I_j$, based on data $n_j$ to be \cite{jack2009}

\begin{equation}
 Ln\mathcal{L}(I_j; n_j)=\sum^N_{j=1} n_j Ln(I_j + \varepsilon) - (I_j + \varepsilon) - Ln(n_j !).
\end{equation}

In the absence of any additional knowledge, the reconstructed image is simply the recorded data itself i.~e.~$I_j = n_j$.  However, given that this data is subject to Poissonian noise it is reasonable to select an image from a large range of statistically plausible alternatives.  Within this range we choose to select the image which has the sparsest discrete cosine transform (DCT).   By defining the coefficients of the spatial frequencies of the whole image as $a_i$ we can define a measure of sparsity through the number of participating spatial frequencies, $DCT_p$, as

\begin{equation}
DCT_{p}(I_j)= \frac {(\sum|a_i|)^2}{\sum|a_i|^2}.
\end{equation}

In our work, this optimisation for $I_j$ is based upon an iterative maximisation of a merit function, $\mathcal{M}$, which combines the log likelihood of the reconstructed image and the participation function of its spatial frequencies as, 

\begin{equation}
\mathcal{M}= Ln\mathcal{L}(I_j,n_j) - \lambda \times DCT_p(I_j).
\end{equation}

\begin{figure}[h!]
\begin{center}
\includegraphics[width=\linewidth]{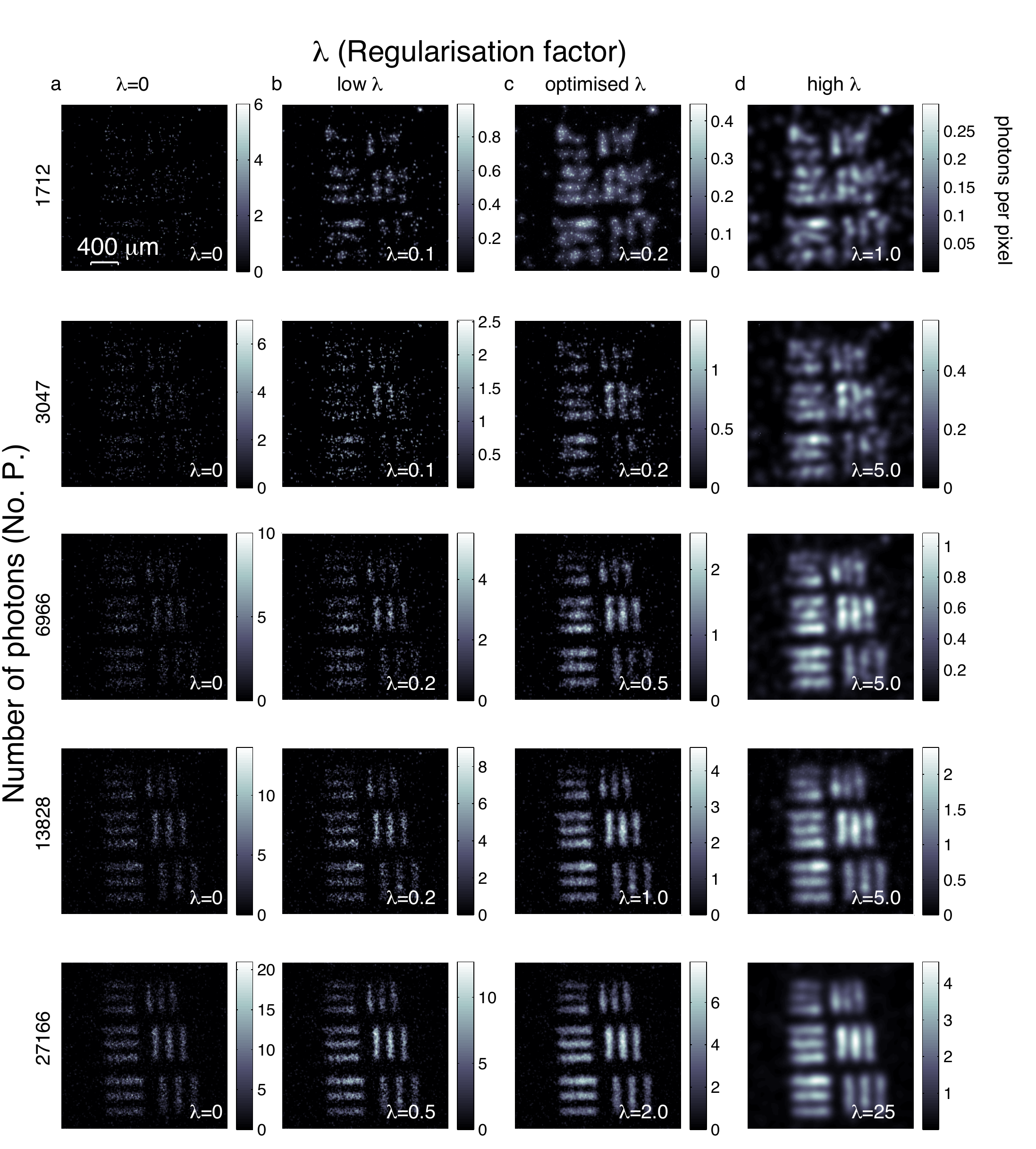}
\caption{Original data in left hand column and reconstructed images for increasing values of $\lambda$ in columns b-d. Column b shows the reconstructed images weighted towards maximising the log likelihood, column d shows the reconstructed images obtained when the optimisation algorithm is overly weighted towards increasing the sparsity in the spatial frequency space and column c shows the reconstructed images with lambda adjusted to give subjectively the best images.}
\label{FIG:TestTargets}
\end{center}
\end{figure}

$\lambda$ is a regularsation factor that sets the balance between a solution that satisfies the recorded data and a solution that satisfies the sparsity condition. Each iteration of our optimisation routine
 makes a random change to the intensity value, $I_j$, of a pixel selected at random. The merit function is calculated for this modified image, and repeated iterations are performed until the image corresponding to a maximisation of this merit function is found. If $\lambda$ is set to zero, the reconstructed image corresponds exactly to the data recorded, whereas if $\lambda$ is set to a very high level, the reconstructed image corresponds to a uniform intensity distribution.

We use our imaging system in the GI configuration, as shown in figure \ref{FIG:setup}-b, where the object, the USAF test target, is placed in the heralding arm of the system and the photons detected by the heralding detector are used to trigger the ICCD camera. We acquire images based on the accumulation of an increasing number of frames and hence of an increasing number of photons and optimise each image using varying values of $\lambda$. Due to the point spread function of the intensifier in the ICCD, the observed resolution of the images is lower than the pixel size on the CCD.  To better match the resolving power of our system to the pixel size in our reconstructed image, we spatially sum our image over adjacent pixels, such that the $600\times600$ pixels of the CCD are processed as a $300\times300$ image.  

\begin{figure}[h!]
\begin{center}
\includegraphics[width=\linewidth]{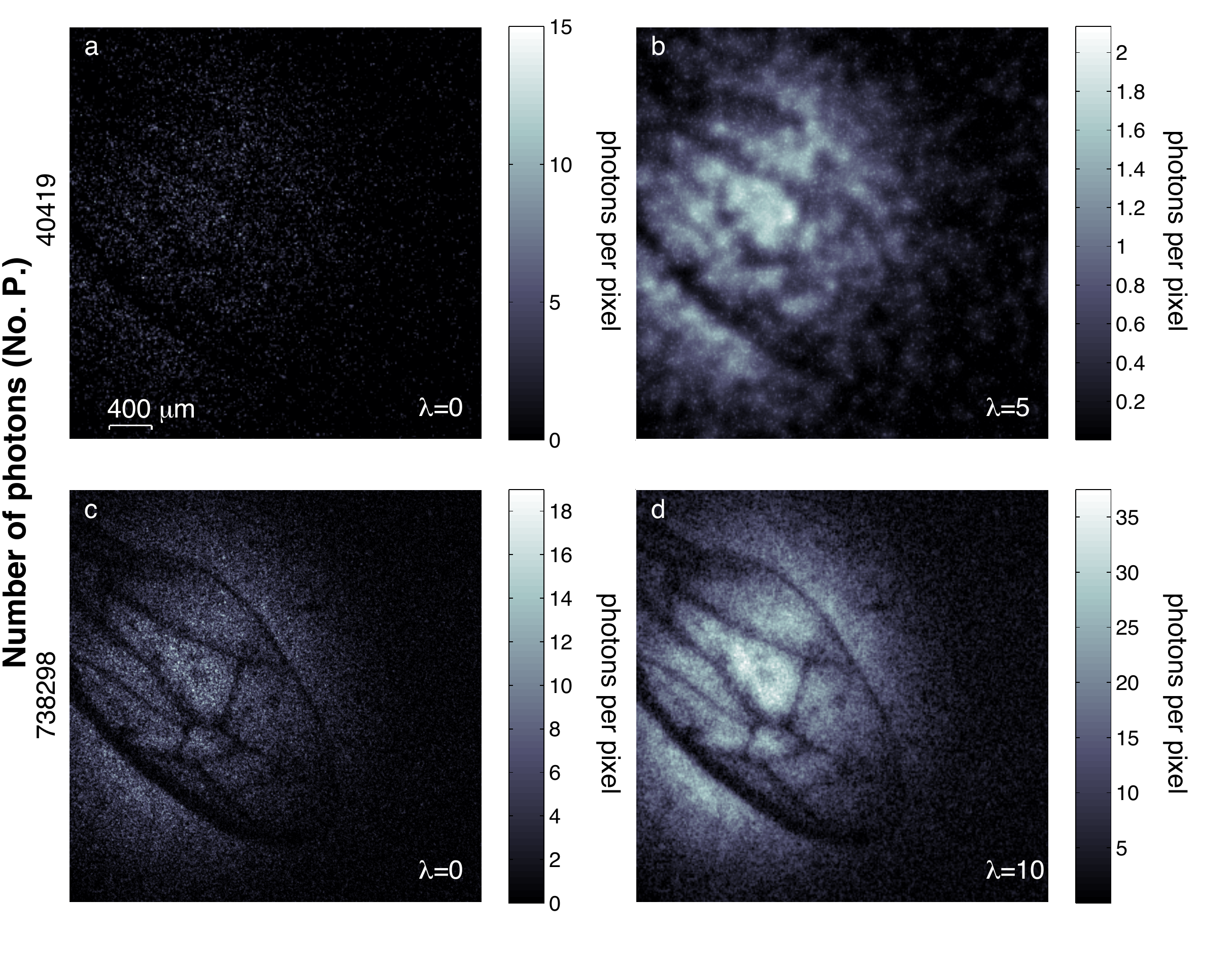}
\caption{How many photons does it take to form an image? a) A weakly absorbing wasp wing imaged using 40419 detected photons and b) the corresponding reconstructed image. c) An image of the same wasp wing with a greater number of photons and d) its associated reconstructed image.}
\label{FIG:WaspWing}
\end{center}
\end{figure}

The reconstructed images, shown in figure \ref{FIG:TestTargets}, highlight the trade-off between changing the relative weighting between the log likelihood of the reconstructed image and participation of spatial-frequencies within the merit function. As $\lambda$ increases, the image becomes smoother due to increasing sparsity in the spatial frequency domain, but for high values of $\lambda$ the resolution is degraded. Figure \ref{FIG:TestTargets} shows the original data, and images for a low-value, optimum-value and high-value of $\lambda$. The lower values of $\lambda$ give images that retain the sparse characteristics of the original data, whereas the high values of $\lambda$ give overly smooth images with associated loss of fine structure. The weighting factors for the central values of $\lambda$ give subjectively the best images.  We see that we are able to form a reconstructed image of the test target for $<7000$ photons, which corresponds to less than $0.2$ photons per image pixel. 

Having established that the system can be used in conjunction with a reconstruction technique to produce images from low numbers of photons we apply the system to the imaging of a biological sample, in our case the wing of a household wasp. The data from this wasp wing for both low and high photon number acquisititions along with their reconstructed images are shown in figure \ref{FIG:WaspWing}.  The low photon number image comprises of of only 40419 detected photons over a field of view of 90,000 image pixels, corresponding to $0.45$ photons per pixel.

\section{Conclusions}

For certain imaging applications a low photon flux is essential, for instance in covert-imaging and biological-imaging where a high photon flux would have detrimental effects. We have developed a low-light imaging technique using a camera-enabled, time-gated imaging system. We exploit the natural sparsity in the spatial frequency domain of typical images and the Poissonian nature of our acquired data to apply image enhancement techniques that subjectively improve the quality of our images. We show that it is possible to retrieve an image of a USAF test target using just 7000 detected photons. These image enhancement techniques, combined with our photon-counting, low-light imaging system, enable the reconstruction of images with a photon number less than one photon per pixel. As an example of low intensity imaging of biological samples, we use this time-gated ghost imaging configuration to acquire low photon number images of a wasp wing, with an average photon-per-pixel ratio of $0.45$.

\begin{acknowledgments}

We acknowledge the financial support from the UK EPSRC under a Program Grant, COAM and the ERC under an Advanced Investigator Grant, TWISTS. R.~W.~B.~acknowledges support form the Canada Excellence Research Chairs programme.
\end{acknowledgments}


\section*{References}
\begin{thebibliography}{10}

\bibitem{strekalov1995}
D.~V. Strekalov, A.~V. Sergienko, D.~N. Klyshko, and Y.~H. Shih.
\newblock Observation of two-photon``ghost'' interference and diffraction.
\newblock {\em Phys. Rev. Lett.}, 74(18):3600--3603, May 1995.

\bibitem{walborn2006}
S.~P. Walborn, D.~S. Lemelle, M.~P. Almeida, and P.~H.~Souto Ribeiro.
\newblock Quantum key distribution with higher-order alphabets using spatially
  encoded qudits.
\newblock {\em Phys. Rev. Lett.}, 96(9):090501, 2006.

\bibitem{dixon2012}
P.~B Dixon, G.~A. Howland, J. Schneeloch, and J.~C. Howell.
\newblock Quantum mutual information capacity for high-dimensional entangled
  states.
\newblock {\em Phys. Rev. Lett.}, 108:143603, 2012.

\bibitem{walborn2010}
S.~P. Walborn, C.~H. Monken, S.~Padua, and P.~H. Souto~Ribeiro.
\newblock Spatial correlations in parametric down-conversion.
\newblock {\em Phys. Rep.}, 495(4-5):87--139, October 2010.

\bibitem{pittman1995}
T.~B. Pittman, Y.~H. Shih, D.~V. Strekalov, and A.~V. Sergienko.
\newblock Optical imaging by means of two-photon quantum entanglement.
\newblock {\em Phys. Rev. A}, 52(5):R3429--R3432, Nov 1995.

\bibitem{lemos2014}
G.~B. Lemos, V.~Borish, G.~D. Cole, S.~Ramelow, R.~Lapkiewicz, and
  A.~Zeilinger.
\newblock Quantum imaging with undetected photons.
\newblock {\em arXiv preprint arXiv:1401.4318}, 2014.

\bibitem{gatti2003}
A.~Gatti, E.~Brambilla, and L.~A. Lugiato.
\newblock Entangled imaging and wave-particle duality: from the microscopic to
  the macroscopic realm.
\newblock {\em Phys. Rev. Lett.}, 90(13), 2003.

\bibitem{bennink2004}
R.~S. Bennink, S.~J. Bentley, R.~W. Boyd, and J.~C. Howell.
\newblock Quantum and classical coincidence imaging.
\newblock {\em Phys. Rev. Lett.}, 92(3):033601, Jan 2004.

\bibitem{edgar2012}
M.~P. Edgar, D.~S. Tasca, F.~Izdebski, R.~E. Warburton, J.~Leach, M.~Agnew,
  G.~S. Buller, R.~W. Boyd, and M.~J. Padgett.
\newblock Imaging high-dimensional spatial entanglement with a camera.
\newblock {\em Nat. Commun.}, 3:984, 2012.

\bibitem{brida2010}
G.~Brida, M.~Genovese, and I.~R. Berchera.
\newblock Experimental realization of sub-shot-noise quantum imaging.
\newblock {\em Nature Photon.}, 4(4):227--230, 2010.

\bibitem{fickler2013}
R.~Fickler, M.~Krenn, R.~Lapkiewicz, S.~Ramelow, and A.~Zeilinger.
\newblock Real-time imaging of quantum entanglement.
\newblock {\em Sci. Rep.}, 3, 2013.

\bibitem{aspden2013}
R.~S. Aspden, D.~S. Tasca, R.~W. Boyd, and M.~J. Padgett.
\newblock {E}{P}{R}-based ghost imaging using a single-photon-sensitive camera.
\newblock {\em New J. Phys.}, 15(7):073032, 2013.

\bibitem{tasca2013}
D.~S. Tasca, R.~S. Aspden, P.~A. Morris, G.~Anderson, R.~W. Boyd, and M.~J.
  Padgett.
\newblock The influence of non-imaging detector design on heralded
  ghost-imaging and ghost-diffraction examined using a triggered {ICCD} camera.
\newblock {\em Opt. Express}, 21(25):30460--73, Dec 2013.

\bibitem{shapiro2012}
J.~H. Shapiro and R.~W. Boyd.
\newblock The physics of ghost imaging.
\newblock {\em Quantum Inf. Process.}, 11(4):949--993, 2012.

\bibitem{morris1984}
G.~M. Morris.
\newblock Image correlation at low light levels: a computer simulation.
\newblock {\em Appl. optics}, 23.18(3152-3159.), 1984.

\bibitem{candes2008}
E.~J. Cand{\`e}s and M.~B. Wakin.
\newblock An introduction to compressive sampling.
\newblock {\em IEEE Signal Proc. Mag.}, 25(2):21--30, 2008.

\bibitem{romberg2008}
J.~Romberg.
\newblock Imaging via compressive sampling [introduction to compressive
  sampling and recovery via convex programming].
\newblock {\em IEEE Signal Proc. Mag.}, 25(2):14--20, 2008.

\bibitem{donoho2006}
D.~L. Donoho.
\newblock Compressed sensing.
\newblock {\em IEEE T. Inform. Theory}, 52(4):1289--1306, 2006.

\bibitem{shechtman2011}
Y. Shechtman, Y.~C Eldar, A. Szameit, and M. Segev.
\newblock Sparsity based sub-wavelength imaging with partially incoherent light
  via quadratic compressed sensing.
\newblock {\em Opt. Express}, 19(16):14807--22, Aug 2011.

\bibitem{shabani2011}
A.~Shabani, R.~L. Kosut, M.~Mohseni, H.~Rabitz, M.~A. Broome, M.~P. Almeida,
  A.~Fedrizzi, and A.~G. White.
\newblock Efficient measurement of quantum dynamics via compressive sensing.
\newblock {\em Phys. Rev. Lett.}, 106(10):100401, 2011.

\bibitem{zerom2011}
P.~Zerom, K.~W.~C. Chan, J.~C. Howell, and R.~W. Boyd.
\newblock Entangled-photon compressive ghost imaging.
\newblock {\em Phys. Rev. A}, 84(6):061804, 2011.

\bibitem{magana2013}
O.~S. Maga{\~n}a-Loaiza, G.~A. Howland, M.~Malik, J.~C. Howell, and R.~W. Boyd.
\newblock Compressive object tracking using entangled photons.
\newblock {\em Appl. Phys. Lett.}, 102(23):231104, 2013.

\bibitem{jack2009}
B.~Jack, J.~Leach, J.~Romero, S.~Franke-Arnold, M.~Ritsch-Marte, S.~M. Barnett,
  and M.~J. Padgett.
\newblock Holographic ghost imaging and the violation of a {B}ell inequality.
\newblock {\em Phys. Rev. Lett.}, 103(8):083602, Aug 2009.

\end{thebibliography}
\end{document}